\documentstyle[aps,epsfig]{revtex} 
\begin{document} 
\draft
\preprint{gr-qc/yymmxxx} 
\title{\large {\bf Logarithmic correction to the Bekenstein-Hawking entropy}}
\author{Romesh K. Kaul and Parthasarathi 
Majumdar\footnote{email: kaul, partha@imsc.ernet.in}}
\address{\it The Institute of Mathematical Sciences, Chennai 600 113, India.}
\maketitle
\begin{abstract}
The exact formula derived by us earlier for the entropy of a four 
dimensional non-rotating black hole within the quantum geometry formulation 
of the event horizon in terms of boundary states of a three dimensional 
Chern-Simons theory, is reexamined for large horizon areas. In addition to
the 
{\it semiclassical} Bekenstein-Hawking contribution proportional to the 
area obtained earlier, we find a contribution proportional to the logarithm of
the area together with subleading corrections that constitute a series in 
inverse powers of the area.  
\end{abstract}

The derivation of the Bekenstein-Hawking (BH) area law for black hole
entropy from the quantum geometry
approach \cite{ash} (and also earlier from string theory \cite{stro} for
some
special cases), has lead to a resurgence of interest in the quantum aspects of
black hole physics in recent times.  However, the major activity has remained
focussed on {\it confirming} the area law for large black holes, which, as
is well-known, was obtained
originally on the basis of arguments of a semiclassical nature. The question
arises as to whether any essential feature of the bona fide quantum aspect of
gravity, beyond the domain of the semiclassical approximation, has been captured
in these assays. Indeed, as has been most eloquently demonstrated by Carlip
\cite{car}, a derivation of the area law alone seems to be possible on the basis
of some symmetry principle of the (semi)classical theory itself without
requiring a detailed knowledge of the actual quantum states associated with a
black hole. The result seems to hold for arbitrary number of spatial dimensions,
so long as a particular set of isometries of the metric is respected. That
quantum gravity has a description in terms of spin networks (or for that matter,
in terms of string states in a fixed background) appears to be of little
consequence in obtaining the area law, although these proposed underlying
structures also lead to the same behaviour via alternative routes, in the
semiclassical limit of arbitrarily large horizon area. 

Although there is as yet no complete quantum theory of gravitation, one
would in general expect key features uncovered so far to lead to
modifications of the area law which could not have been anticipated through
semiclassical reasoning. Thus, the question as to what is the dominant quantum
correction due to these features of quantum gravity becomes one of paramount
importance. Already in the string theory literature \cite{wit} examples of
leading corrections to the area law, obtained by counting D-brane states
describing special supersymmetric extremal black holes (interacting with
massless vector supermultiplets) have appeared. This has received strong support
recently from semiclassical calculations in $N=2$ supergravity \cite{dew}
supplemented by ostensible stringy higher derivative corrections which are
incorporated using Wald's general formalism describing black hole entropy as
Noether charge \cite{wald}. However, the geometrical interpretation of
these corrections remains unclear. Further, there are subtleties
associated with direct application of Wald's formalism which assumes a
{\it non-degenerate} bifurcate Killing horizon, to the case of
extremal black holes which have degenerate horizons. Moreover, the string
results do not pertain to generic (i.e., non-extremal) black holes of
Einstein's general relativity, and are constrained by the unphysical
requirement of unbroken spacetime supersymmetry. 

In this paper, we consider the corrections to the semiclassical area law
of generic four dimensional non-rotating black holes, due to key aspects
of {\it non-perturbative} quantum gravity (or quantum geometry) 
formulated by Ashtekar and collaborators \cite{ash2}. In \cite{ash},
appropriate boundary conditions are imposed on dynamical variables at the
event horizon considered as an inner boundary. These boundary conditions
require that the Einstein-Hilbert action be supplemented by boundary terms
describing a three dimensional $SU(2)$ Chern-Simons theory living on a
finite `patch' of the horizon with a spherical boundary, punctured by
links of the spin network bulk states describing the quantum spacetime
geometry interpolating between asymptopia and the horizon. On this two
dimensional boundary there exists an $SU(2)$ Wess Zumino model whose
conformal blocks describe the Hilbert space of the Chern-Simons theory
modelling the horizon. An exact formula for the number of these conformal
blocks has been obtained by us \cite{km} two years ago, for arbitrary level
$k$ and number of punctures $p$. It has been shown that in the limit of large
horizon area given by arbitrarily large $k$ and $p$, the logarithm of
this number duly yields the area law. Here we go one step further, and
calculate the dominant sub-leading contribution, as a function of the
classical horizon area, or what is equivalent, as a function of the BH
entropy itself. 

On purely dimensional grounds, one would expect the entropy to have an
expansion, for large classical horizon area, in inverse powers of area so
that the BH term is the leading one,
\begin{equation}
S_{bh}~~=~~S_{BH}~+~\sum_{n=0}^{\infty} C_n~A_H^{-n}~~\label{gen}
\end{equation}
where, $A_H$ is the classical horizon area and $C_n$ are coefficients
which are independent of the horizon area but dependent on the Planck
length (Newton constant). Here the Barbero-Immirzi parameter \cite{barb}  
has been `fitted' to the value which fixes the normalization of the BH term
to the standard one. However, in principle, one could expect an additional
term proportional to $ln~A_H$ as the leading quantum correction to the
semiclassical $S_{BH}$. Such a term is expected on general grounds
pertaining to breakdown of na\"ive dimensional analysis due to quantum
fluctuations, as is common in quantum field theories in flat spacetime
and also in quantum theories of critical phenomena. We show, in what
follows, that such a logarithmic correction to the semiclassical area law
does indeed arise from the formula derived earlier \cite{km} and derive
its coefficient. 

We first briefly recapitulate the derivation \cite{km} of the general
formula for the number of conformal blocks of the $SU(2)_k$ Wess Zumino
model on a punctured 2-sphere appropriate to the black hole situation. This
number can be computed in terms of the so-called fusion matrices
$N_{ij}^{~~r}$ \cite{dms} 
\begin{equation} 
N_{\cal P}~=~~\sum_{\{r_i\}}~N_{j_1 j_2}^{~~~~r_1}~ N_{r_1 j_3}^{~~~~r_2}~
N_{r_2 j_4}^{~~~~r_3}~\dots \dots~ N_{r_{p-2} j_{p-1}}^{~~~~~~~~j_p} ~
\label{fun} \end{equation} 
Diagrammatically, this can be represented as shown in fig. 1 below. 
\begin{figure}[htb] \begin{center}
\mbox{\epsfig{file=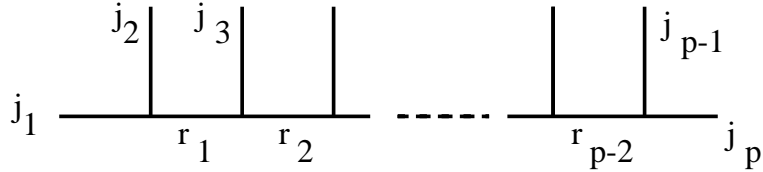,width=10truecm,angle=0}} 
\caption{Diagrammatic representation of composition of spins $j_i$ for
$SU(2)_k$} 
\end{center}
\end{figure} 
Here, each matrix element $N_{ij}^{~~r}$ is $1 ~or~ 0$,
depending on whether the primary field $[\phi_r]$ is allowed or not in the
conformal field theory fusion algebra for the primary fields $[\phi_i]$
and $[\phi_j] $ ~~($i,j,r~ =~ 0, 1/2, 1, ....k/2$): 
\begin{equation}
[\phi_i] ~ \otimes~ [\phi_j]~=~~\sum_r~N_{ij}^{~~r} [\phi_r]~ . 
\label{fusal}
\end{equation}
Eq. (\ref {fun}) gives the number of conformal blocks with spins $j_1,
j_2, \dots, j_p$ on $p$ external lines and spins $r_1, r_2, \dots,
r_{p-2}$ on the internal lines. 

We then use the Verlinde formula \cite{dms} to obtain
\begin{equation}
N_{ij}^{~~r}~=~\sum_s~{{S_{is} S_{js} S_s^{\dagger r }} \over S_{0s}}~,
\label{verl} 
\end{equation} 
where, the unitary matrix $S_{ij}$ diagonalizes the fusion matrix. Upon
using the unitarity of the $S$-matrix, the algebra (\ref{fun}) reduces to 
\begin{equation} 
N_{\cal P}~=~ \sum_{r=0}^{k/2}~{{S_{j_1~r} S_{j_2~r} \dots S_{j_p~r}}
\over (S_{0r})^{p-2}}~. 
\label{red} \end{equation} 
Now, the matrix elements of $S_{ij}$ are known for the case under
consideration ($SU(2)_k$ Wess-Zumino model); they are given by
\begin{equation} 
S_{ij}~=~\sqrt{\frac2{k+2}}~sin \left({{(2i+1)(2j+1) \pi} \over k+2}
\right )~, \label{smatr}
\end{equation} 
where, $i,~j$ are the spin labels, $i,~j ~=~ 0, 1/2, 1, .... k/2$. Using
this $S$-matrix, the number of conformal blocks for the set of punctures
${\cal P}$ is given by
\begin{equation} 
N_{\cal P}~=~{2 \over {k+2}}~\sum_{r=0}^{ k/2}~{
{\prod_{l=1}^p sin \left( {{(2j_l+1)(2r+1) \pi}\over k+2} \right) } \over
{\left[ sin \left( {(2r+1)  \pi \over k+2} \right)\right]^{p-2} }} ~. 
\label{enpi} \end{equation} 
Eq. (\ref{enpi}) thus gives the dimensionality of the $SU(2)$ Chern-Simons
states corresponding to a three-fold bounded by a two-sphere punctured at
$p$ points. The black hole microstates are counted by summing $N_{\cal P}$
over all sets of punctures ${\cal P}$, $N_{bh}=\sum_{\{\cal P\}} N_{\cal
P}$. Then, the entropy of the black hole is given by $S_{bh}~=~\log
N_{bh}$. 

We are however interested only in the leading correction to the semiclassical
entropy which ensues in the limit of arbitrarily large $A_H$. To this end,
recall that the eigenvalues of the area operator \cite{ash2} are given by
\begin{equation}
A_H~=~8\pi \beta ~l_{Pl}^2~\sum_{l=1}^p~[j_l(j_l+1)]^{\frac12}~,
\label{area} \end{equation}
where, $l_{Pl}$ is the Planck length, $j_l$ is the spin on the $l$th
puncture on the 2-sphere and $\beta$ is the Barbero-Immirzi parameter
\cite{barb}. Clearly, the large area limit corresponds to the limits $k
~\rightarrow~\infty~,~p~\rightarrow~\infty$. Now, from eq. (\ref{area}), it
follows that the number of punctures $p$ is largest for a given $A_H$
provided {\it all} spins $j_l~=~\frac12$. Thus, for a fixed classical
horizon area, we obtain the largest number of punctures $p_0$ as
\begin{equation}
p_0~=~{A_H \over 4 l_{Pl}^2}~{\beta_0 \over \beta}~, \label{pmax}
\end{equation}
where, $\beta_0=1/\pi \sqrt{3}$. In this approximation, the set of punctures
${\cal P}_0$ with all spins equal to one-half dominates over all other
sets, so that the black hole entropy is simply given by 
\begin{equation}
S_{bh}~~=~~ln~N_{{\cal P}_0}~~, \label{ent}
\end{equation}
with $N_{{\cal P}_0}$ being given by eq. (\ref{enpi}) with $j_l =
1/2$. 

Observe that $N_{{\cal P}_0}$ can now be written as 
\begin{equation}
N_{{\cal P}_0}~=~{2^{p_0+2} \over k+2}~\left[
F(k,p_o)~-~F(k,p_0+2)
\right]~\label{hlf} \end{equation}
where, 
\begin{equation}
F(k,p)~=~\sum^{[\frac12(k+1)]}_{\nu=1}~cos ^p \left({\nu \pi \over k+2}
\right)~. \label{eff} \end{equation}
The sum over $\nu$ in eq. (\ref{eff}) can be approximated by an integral
in the limit $k~\rightarrow~\infty~,~p_0~\rightarrow~\infty$, with
appropriate care being taken to restrict the domain of integration; one
obtains
\begin{equation}
F(k,p_0)~\approx~\left({k+2 \over \pi}
\right)~\int_0^{\pi/2}~dx~cos^{p_0} x~, \label{inte} \end{equation}
so that,
\begin{equation}
N_{{\cal P}_0}~\approx~{2^{p_0+2} \over \pi (p_0 +2)}~B~({p_0+1
\over 2}~,~\frac12)~, \label{nap} 
\end{equation}
where, $B(x,y)$ is the standard $B$-function \cite{whit}. Using well-known
properties of this function, it is straightforward to show that
\begin{eqnarray}
ln~N_{{\cal P}_0}&~=~&
p_0~ln2~-~\frac32~ln~p_0~-~ln~(2\pi)~\nonumber \\
&-&~\frac52~p_0^{-1}~+~O(p_0^{-2})~~. \label{lnn}
\end{eqnarray}
Substituting for $p_0$ as a function of $A_H$ from eq. (\ref{pmax}) and
setting the Barbero-Immirzi parameter $\beta$ to the `universal' value
$ \beta_0~ln2 $ \cite{ash}, one obtains our main result
\begin{equation}
S_{bh}~=~S_{BH}~-~{3\over 2}~ln\left( S_{BH} \over ln2
\right)~+~const.~+~\cdots~, \label{main}
\end{equation}
where, $S_{BH}=A_H/4 l_{Pl}^2$, and the ellipses denote corrections in
inverse powers of $A_H$ or $S_{BH}$. 

Admittedly, the above calculation is restricted to the leading correction
to the semiclassical approximation. It has been done for a fixed large
$A_H$ by taking the spins on all the punctures to be 1/2 so that we have
the largest number of punctures. But it is not difficult to argue that
the coefficient of the $ln A_H$ term is robust in that inclusion of spin
values higher than 1/2 do not affect it, although the constant term and
the coefficients of sub-leading corrections with powers of $O(A_H^{-1})$
might get affected. The same appears to be true for values of
the level $k$ away from the asymptotic value which we have assigned it
above: the coefficient of the $ln A_H$ is once again unaffected. Thus, the
leading logarithmic correction with coefficient -3/2 that we have
discerned for the black hole entropy is in this sense {\it universal}. 
Moreover, although we have set $\beta=\beta_0~ln2$ in the above formulae,
the coefficient of the $ln A_H$ term is independent of $\beta$, a feature
not shared by the semiclassical area law. 

It is therefore clear that the leading correction (and maybe also the
subleading ones) to the BH entropy is negative. One way to understand this
could be the information-theoretic approach of Bekenstein \cite{bek}: black
hole entropy represents lack of information about quantum states which arise
in the various ways of gravitational collapse that lead to formation of black
holes with the same mass, charge and angular momentum. Thus, the BH entropy
is the `maximal' entropy that a black hole can have; incorporation of
leading quantum effects reduces the entropy. The logarithmic nature of
the leading correction points to a possible existence of what might be
called a `non-perturbative fixed point'. That this happens in the physical
world of four dimensions is perhaps not without interest. 

Recently, the zeroth and first law of black hole mechanics have been
derived for situations with radiation present in the vicinity
of the horizon, using the notion of the isolated horizon \cite{ash3}. Our
conclusions above for the case of non-rotating black holes hold for such
generalizations \cite{ash4} as well. Note however that while, 
the foregoing analysis involves $SU(2)_k$ Chern Simons theory, for large 
$k$ this reduces to a specific $U(1)$ theory presumably related to 
the `gauge fixed' classical theory discussed in \cite{ash3}. The charge 
spectrum of this $U(1)$ theory is discrete and bounded from 
above by $k$. The $SU(2)$ origin of the theory thus provides a natural 
`regularization' for calculation of the number of conformal blocks. 

\noindent {\it Note Added:} After the first version of this paper appeared
in the Archives, it has been brought to our attention that corrections to
the area law in the form of logarithm of horizon area have been obtained
earlier \cite{mann} for extremal Reissner-Nordstrom and dilatonic black
holes. These corrections are due to quantum scalar fields propagating in
fixed classical backgrounds appropriate to these black holes. The
coefficient of the $ln A_H$ term that appears in ref. \cite{mann} is
different from ours. This is only expected, since in contrast to ref.
\cite{mann}, our corrections originate from non-perturbative quantum
fluctuations of spacetime geometry (for generic non-rotating black holes),
{\it in the absence of matter fields.} Thus, this correction is {\it
finite} and independent of any arbitrary `renormalization scale'
associated with divergences due to quantum matter fluctuations in a fixed 
classical background. 

We thank Prof. A. Ashtekar for many illuminating discussions and Prof. R.
Mann for useful correspondence.

\end{document}